

\magnification \magstep1
\openup 1.1\jot
\input epsf
\font\reffont=cmcsc10

\line{July 1994. Revised Mar 1995. \hfill DAMTP/94-39; hep-th/9501117}
\vskip 50pt

\centerline {{\bf A Stochastic Model of a
              Quantum Field Theory}\footnote*{To appear in J. Stat. Phys.}
              }
\vskip 40pt

\centerline {\reffont T.M. Samols}
\vskip 15pt

\centerline {\it Department of Applied Mathematics and Theoretical Physics,}
\centerline {\it University of Cambridge, Silver Street, Cambridge CB3 9EW,}
\centerline {\it United Kingdom}
\centerline {and}
\centerline {\it King's College,}
\centerline {\it Cambridge CB2 1ST,}
\centerline {\it United Kingdom}
\vskip 40pt

\centerline{\bf Abstract}
\vskip 5pt
The problem of obtaining a realistic, relativistic
description of a quantum system is discussed in the context of
a simple (light-cone) lattice field theory. A natural stochastic
model is proposed which, although non-local, is relativistic
(in the appropriate lattice sense), and which is operationally
indistinguishable from the standard quantum theory. The
generalization to a broad class of lattice theories is briefly
described.
\vskip 20pt
\centerline{\bf 1. Introduction}
\vskip 5pt
Quantum theory is a highly successful algorithm for predicting
the results of experiments. It is, however, beset with
worrying conceptual problems. These may be traced to the
fact that although one may calculate
probabilities, there is no objectively
defined space of events to which these probabilities refer. In
a typical interpretation, a space of events must first be
chosen from an infinite number of incompatible alternatives, and
probabilities may only be extracted after this choice is made.
The choice -- whether it is effected by means of a division of the world
into system and observer, or system and environment; or whether it is a
choice of a subalgebra of observables, or one among many sets of
consistent histories -- is still in the end a subjective choice.
Its role seems inappropriate in what
is supposed to be a fundamental theory, and to beg the
question of how to explain the very particular space of events
that constitutes the world of our actual experience, namely the
classical one.$^{(1)}$
\vskip 5pt
To a realist, the form that a remedy must take is clear:
a space of objectively defined events must be restored to the theory.
If difficulties are encountered in pursuing this course, then they
will be entirely conventional ones -- those of meeting the demands
of predictive power, simplicity, consistency, and so on -- but
understanding the relationship of the theory to the experience
it is meant to describe should not pose special philosophical
problems.
\vskip 5pt
Realistic theories have essentially been of
two types: either the wavefunction evolves according to the standard
unitary law and is supplemented by some further
variables with prescribed dynamics, as for instance in the theory of
de Broglie and Bohm$^{(2)}$, and Bell's stochastic generalization
to quantum field theory$^{(3,4)}$;
or alternatively, a definite departure from quantum theory is
contemplated and the wavefunction is subject to a stochastic
and non-unitary evolution (see e.g. Pearle$^{(5)}$, Percival$^{(6)}$
and refs. therein).
Of particular practical interest is that theories of the second type,
by predicting deviations from standard results, are beginning to
suggest experimental tests of quantum theory itself.
\vskip 5pt
As is well known from the work of Bell$^{(1,7)}$ realistic theories must
inevitably be non-local in character. The purpose here is to
examine the question of whether, despite this non-locality, it
is possible to preserve relativistic invariance.
More precisely, we would like a theory that is ``fundamentally''
relativistic: not only should it enjoy
phenomenological Lorentz invariance (as, for instance, the
theory of Lorentz himself), but in addition, its
formulation should not rest on the choice of a
preferred frame.
\vskip 5pt
The proper setting for this question is quantum field theory and
the discussion will be based on a particularly simple example
-- a light-cone lattice field theory in one space dimension --
though as we remark at the end, the generalization to a large
class of theories is straightforward.
On the lattice of course one cannot have full Lorentz symmetry.
However, there is a causal structure and
the notion of a spacelike surface, and in this restricted context
a ``relativistic theory'' shall mean one in which the dynamics
does not depend on a preferred choice of such surfaces. For an
appropriate lattice theory one expects to recover full Lorentz
invariance in the continuum limit.
\vskip 5pt
We shall show that the most
na\"{\i}ve attempt at a realistic formulation
fails, and then propose a rather natural model with the desired properties.
The model is of the first type mentioned above:
thus the mathematical structure of quantum theory is left
completely intact, but is supplemented by extra variables governed
by a stochastic evolution law.
It is very much an attempt at a minimal
solution to the problem and agreement with the results of
conventional quantum theory is built in, although one might
also regard it as the starting point for a theory that differs from
the conventional theory in a testable way.
Perhaps of particular interest is the generality of the construction:
essentially the only features of the underlying quantum theory
that it requires are the causal structure and the local,
unitary evolution law. One may thus obtain a straightforward
probabilistic description of a rather general class of local
lattice quantum theories.
\vskip 5pt
For some related ideas, and some similar points regarding measurement,
see the discussion of how the Bohm theory might be made relativistic
by D\"urr {\it et al.}$^{(8)}$ and Berndl {\it et al.}$^{(9)}$ (and
also refs. 10, 11). For another approach, in the context
of a theory of the second type, see Ghirardi {\it et al.}$^{(12)}$.
\vskip 10pt
\centerline{\bf 2. The quantum theory}
\vskip 5pt
Light-cone lattice field theory has been introduced in the study of
integrable models in $(1+1)$ dimensions;$^{(13)}$
in statistical mechanics the analogue is the diagonal-to-diagonal
transfer matrix method. It exhibits in a particularly transparent
form the essential elements of any local quantum field theory --
a causal structure, and a local unitary evolution law adapted to that
structure.

$$\displaylines{
 \epsfbox{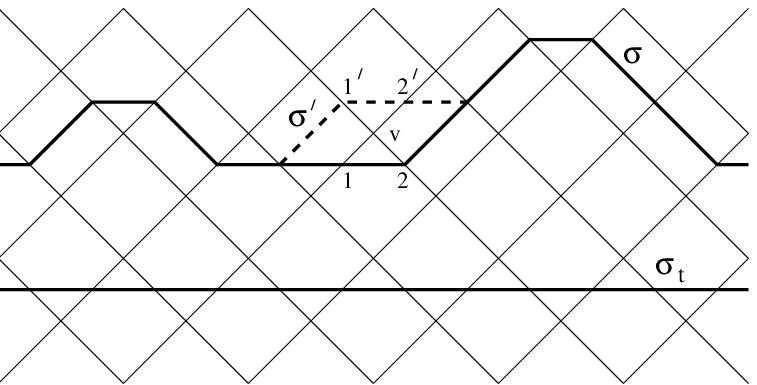}}$$
\centerline{\vbox{\hsize=5.5 true in \noindent {\bf Figure 1.}
   {\it The light-cone
  lattice. $\sigma_t$ is a constant time slice; $\sigma$ is a general
  spacelike surface, and $\sigma'$ one obtained from it by an elementary
  motion.}}}
\vskip 5pt
The spacetime is represented by a lattice generated by null
rays as shown in figure 1, and the local observables
of the theory live on the links. In the simplest theory there are just
two states associated with each link $l$ labelled by
$\alpha_l=0, 1$, which we shall refer to as the occupation number.
At each vertex of the lattice the local evolution law is
encoded in a 4-dimensional unitary matrix
-- an ``$R$-matrix'' -- whose
entries, $R_{\alpha_1 \, \alpha_2}^{\alpha_{1'} \alpha_{2'}}$, are the
amplitudes connecting the four possible states on the ingoing
and outgoing pairs of links at that point
(see figure 2{\it a}).
The causal structure is the obvious one:
two links are spacelike separated (the corresponding
local operators on the links commute) if and only if there is no
everywhere future-directed path on the lattice connecting them.

$$\displaylines{
 \epsfbox{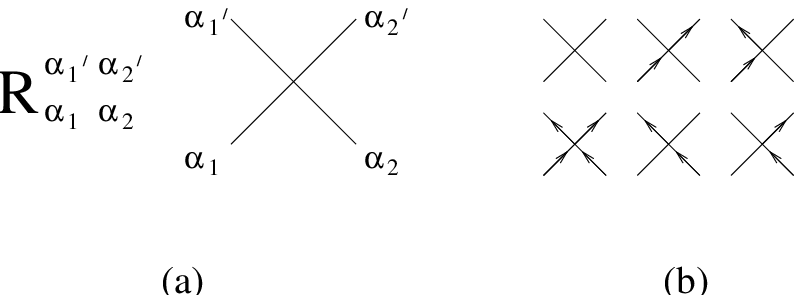} \cr} $$
\centerline{
 \vbox{\hsize=5.5 true in \noindent {\bf Figure 2. (a)}{\it The R-matrix
 associated with a vertex.}
{\bf(b)} {\it A possible set of non-zero amplitudes conserving occupation
number. An arrow on a link denotes occupation number one.}}}
\vskip 5pt
A quantum state $\Psi$ is fully determined by a complex function
(the wavefunction) of
the variables on the links cut by a constant time slice, $\sigma_t$.
Denoting this set of variables by $\alpha|_{\sigma_t}$, we write the
wavefunction as $\Psi(\alpha|_{\sigma_t})$.
The unitary evolution to the wavefunction on the next constant time
slice is then effected by multiplying by all the $R$-matrices
associated with the vertices lying just to the future of $\sigma_t$,
and summing over the repeated $\alpha_l$'s.
\vskip 5pt
More generally,
one may consider the wavefunction $\Psi(\alpha |_{\sigma})$ on any
spacelike surface $\sigma$. To evolve it to another
surface, $\sigma'$, one now applies all the $R$-matrices associated
with the vertices in the region between $\sigma$ and $\sigma'$.
(We assume that $\sigma'$ is everywhere coincident with, or
to the future of $\sigma$, although transformations between intersecting
surfaces can be considered just as well.)
In the simplest case, when the region contains a single
vertex, $v$ say, we shall call the local deformation of
$\sigma$ to $\sigma'$ an ``elementary motion'', and
only one $R$-matrix need be applied.
Thus, if $l=1, 2$ and $1',2'$ are the ingoing and outgoing links
respectively at $v$, as shown in figure 1, the evolution of
the wavefunction is given by

$$ \Psi(\alpha |_{\sigma'}) = \sum_{\alpha_1 \alpha_2}
 R_{\alpha_1 \, \alpha_2}^{\alpha_{1'} \alpha_{2'}}
 \Psi(\alpha |_{\sigma}).  \eqno (1) $$

\noindent
The evolution between two arbitrary surfaces can
be obtained by composing elementary transformations of this type.
To fix the boundary conditions, we take the lattice to be
periodic in space, of width $2N$; the spacetime is then a
(discretized) cylinder and the wavefunction on any surface is a
function of $2N$ variables.
\vskip 5pt
The standard interpretation of the theory is expressed in terms
of the results of measurements of arbitrary hermitian
operators associated with any surface $\sigma$.
It is sufficient, however, to restrict to the projection
operators corresponding to the joint occupation number eigenstates
labelled by $\alpha |_{\sigma}$.
The point of view represented by this restriction is
essentially the familiar one that, ultimately, all
measurements reduce to measurements of position.
The predictions of the theory are then summarized by the rule
that, in a state $\Psi$, the probability $p_\Psi(\alpha |_{\sigma})$
of finding the configuration $\alpha |_{\sigma}$ on the surface
$\sigma$ is given by

$$ p_\Psi(\alpha |_{\sigma}) = |\Psi(\alpha |_{\sigma})|^2. \eqno (2) $$

\noindent
In what follows we shall use the phrase ``standard quantum theory''
to mean the mathematical formalism together with this standard
interpretation, though of course it should be borne
in mind that the latter, relying as it does on the notion of
measurement, is not very precise. Note that for an incomplete set
of spacelike separated variables
$\alpha_{l_1},\ldots,\alpha_{l_n}$ $(n<2N)$, the joint probability
may be calculated as the marginal distribution of (2)
for any $\sigma$ that cuts the corresponding links $l_1,\dots,\l_n$;
that this distribution is independent of the choice of $\sigma$,
and so well-defined, is an immediate consequence of the local
unitary evolution of the wavefunction.
\vskip 5pt
This completes then our description of the lattice quantum field theory.
Note that we have not yet made any particular assumption about the
$R$-matrices, though for a conventional field theory with spacetime
translation invariance they will be
uniform over the lattice. With an appropriate choice, and taking
a suitable continuum limit, one obtains for example the massive
Thirring model. The non-zero amplitudes for this case
are depicted in figure 2{\it b}, where occupation number one is indicated
by a forward-pointing arrow. One may think of such arrows
as the paths of ``bare fermions'' through the lattice, though
these are not to be identified with the
physical particles of the eventual continuum theory, since they
are built on the wrong vacuum. For further details the reader
is encouraged to consult Destri \& De Vega.$^{(13)}$ Here we need only
note that we have a lattice system that is rich enough to yield a
non-trivial quantum field theory in the continuum limit.
\vskip 10pt
\centerline{\bf 3. Realistic framework}
\vskip 5pt
Let us now try to construct a realistic model of the system. From the
point of view of the quantum theory there is nothing particularly
special about the variables $\alpha_l$; transformation theory allows
the choice of many other sets of (generally non-local) variables
just as well. However, for a realistic theory it is natural to take
the $\alpha_l$ as fundamental, and to elevate them, in Bell's
terminology, to the status
of ``beables'' -- in other words, to suppose that in the
time evolution of the system each variable $\alpha_l$ realizes
a definite value $\hat\alpha_l$ with some probability, as part of
an objectively defined physical process.
More precisely, given a state $\Psi$, we suppose that there is an
associated joint probability distribution $p_\Psi(\hat\alpha)$ for
the realized values $\hat\alpha_l$ on the entire spacetime lattice.
(To avoid difficulties of definition of $p_\Psi(\hat\alpha)$, we
may restrict attention to a finite number of variables.
Thus in what follows $p_\Psi(\hat\alpha)$ should be regarded
as the distribution associated with the variables between two
bounding surfaces, with the understanding that these
may be moved arbitrarily far into the past and future respectively.)
\vskip 5pt
To secure agreement with standard quantum theory in this framework,
and avoid any reference to a particular frame, it seems simplest to
require that all the quantum mechanical probabilities (2)
arise as the appropriate marginal distributions of
$p_\Psi(\hat\alpha)$, so that for {\it all} surfaces $\sigma$,

$$  \sum_{\hat\alpha |_{\sigma^c}} p_\Psi(\hat\alpha) =
        |\Psi(\hat\alpha |_{\sigma})|^2,                    \eqno(3) $$

\noindent where $\sigma^c$ denotes
all those links of the lattice not cut by $\sigma$. However, this
seemingly natural procedure is too na\"{\i}ve. Indeed, it is one
of the remarkable properties of quantum theory that in general,
no such $p_\Psi(\hat\alpha)$ can be found.
\vskip 5pt
This result follows immediately from the interpretation of the Bell
inequalities as conditions for the existence of
a joint distribution.\footnote{$^1$}{The equivalent result
in the context of a realistic particle mechanics has also been
shown by Berndl {\it et al.}$^{(9,10)}$ (making use of an inequality
of Hardy$^{(15)}$), and was previously conjectured for a field
theory in ref. 8.}
It suffices to consider the simple arrangement shown in figure 3, with
variables $\alpha_1$, $\alpha_{1'}$, $\alpha_2$, $\alpha_{2'}$ at
just four sites -- by an appropriate choice of
$R$-matrices this may be easily embedded in the lattice field theory.
The site $1'$ lies in the causal future
of $1$, the corresponding amplitudes are summarised in
the unitary matrix $R(1)_{\alpha_1}^{\alpha_{1'}}$, and in a spacelike
separated region we have a similar arrangement for
the 2-variables. There are four spacelike
surfaces that can be drawn through the sites and thus four probability
distributions $p(\hat\alpha_1, \hat\alpha_2)$,
$p(\hat\alpha_1, \hat\alpha_{2'})$, $p(\hat\alpha_{1'}, \hat\alpha_2)$,
and $p(\hat\alpha_{1'}, \hat\alpha_{2'})$
for which the standard quantum rule, (2), provides predictions.
For these distributions to be obtainable
as the marginals of an overall distribution
$p(\hat\alpha_1, \hat\alpha_2, \hat\alpha_{1'}, \hat\alpha_{2'})$,
it is necessary (and in fact also sufficient) that they
satisfy the inequalities$^{(14)}$

$$ -1 \leq p(\hat\alpha_1, \hat\alpha_2) - p(\hat\alpha_1,
1-\hat\alpha_{2'}) - p(\hat\alpha_{1'}, \hat\alpha_2)
- p(1-\hat\alpha_{1'}, \hat\alpha_{2'}) \leq 0.         \eqno(4) $$

\noindent
However, for an appropriate choice of state $\Psi$, and matrices $R(i)$,
the predictions of (2) violate these inequalities.
For instance, take the state $\Psi(\alpha_1, \alpha_2)= {1\over\sqrt 2}
(\delta_{\alpha_1 1}\delta_{\alpha_2 0}- \delta_{\alpha_1 0}
\delta_{\alpha_2 1})$, and let $R(1)=\exp{1\over 2}i\theta\sigma_1$,
and $R(2)=\exp-{1\over 2}i\theta\sigma_1$, where $\sigma_1$ is
the first Pauli matrix. Then with
$\hat\alpha_1=0$, $\hat\alpha_2=0$, $\hat\alpha_{1'}=1$,
$\hat\alpha_{2'}=0$, the
above combination of probabilities is, according to (2),
$-{1\over 2}(2 + \cos\theta - \cos^2\theta)$, which violates the
inequality for $-{1\over 2}\pi<\theta <{1\over 2}\pi$.
This is, of course, the familiar mathematics
of the Einstein-Podolsky-Rosen-Bohm experiment, albeit with a somewhat
different interpretation.

$$\displaylines{
 \epsfbox{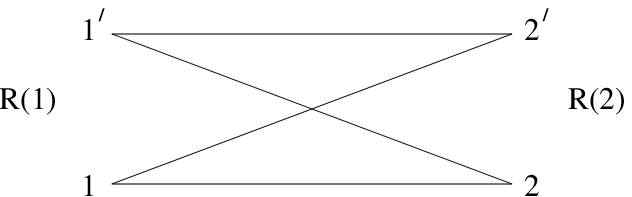} \cr}$$
\centerline{
 \vbox{\hsize=5.5 true in \noindent {\bf Figure 3.} {\it The
impossibility argument: $1'$ is in the future of $1$, and $2'$
in the future of $2$, the two pairs being spacelike separated. The
lines indicate the four spacelike surfaces that can be drawn
through these sites.}}}

\vskip 5pt
In general then, the condition that (3) holds for all surfaces
$\sigma$ must be relaxed. In fact, for operational equivalence
with standard quantum theory, only much weaker conditions are
required -- ones which can be satisfied consistently.
To understand this, consider a general set of spacelike
separated variables $\alpha_{l_1},\ldots,\alpha_{l_n}$. The joint
distribution $p(\hat\alpha_{l_1},\ldots,\hat\alpha_{l_n})$ only
acquires {\it operational}
meaning if the realized values $\hat\alpha_{l_r}$ can be
compared, i.e. if there exist records of these values
that can be brought together to the neighbourhood of the same point
(see ref. 8 for a similar point).
For agreement with quantum theory, we therefore require as a
first condition (I) on $p_\Psi(\hat\alpha)$ that
its {\it local} marginal distributions reproduce the quantum
mechanical results. For our simple lattice theory, these local
distributions are just the two-variable distributions associated
with the ingoing
pairs of links at each vertex. Of course this reasoning presupposes
that the necessary records can in fact be made.
Indeed, rather more generally, we require that
under the appropriate circumstances
(those associated with classical behaviour in standard quantum theory)
continuous, quasi-deterministic trajectories in the appropriate
quantities should emerge. We are therefore led to a second condition (II)
on $p_\Psi(\hat\alpha)$: that it enforce sufficient continuity in
time of the realizations $\hat\alpha_l$ to ensure that
such trajectories appear.
\vskip 5pt
We shall presently remove the imprecision in these heuristic remarks in
the context of our specific model, but before doing so it is useful
to illustrate them with two examples of realistic prescriptions in which
there is a preferred frame and (3) is satisfied {\it only}
on the constant time slices $\sigma_t$. In the first,
one simply picks a configuration on each $\sigma_t$ independently,
in accordance with the distributions $|\Psi(\hat\alpha|_t)|^2$.
This prescription is Bell's (deliberately pathological) single-world
version of the Everett theory.$^{(16)}$ Condition (I) is then
satisfied, but not condition (II); the realizations on each slice
occur with probabilities precisely according to the quantum
prediction, but the independence of the slices means that they
do not fit together to form sensible histories.
The second example is Bell's stochastic generalization of the
Bohm theory$^{(3)}$, formulated for a fermionic field
theory on a spatial lattice, with time
kept continuous. Here, configurations on (infinitesimally close)
time slices are connected by transition probabilities proportional
to the transition amplitudes between the corresponding eigenstates.
This transition rule (a generalization of Bohm's guiding condition)
enforces the necessary continuity in time, so that (II) is satisfied
as well. If one restricts attention to those quantities with
quasi-classical behaviour -- which is the level at which the results
of all measurements must be recorded -- this is enough to ensure
agreement (for these quantities) with the quantum results on all,
and not just constant time surfaces.
One thus has a form of relativistic invariance, but it is
phenomenological rather than fundamental. Our aim here is to
go one step further and provide a scheme that is
fully relativistic.
\vskip 10pt
\centerline{\bf 4. The stochastic model}
\vskip 5pt
Let the system be in the state $\Psi$. We describe a
simple stochastic model which generates a joint distribution
$p_\Psi(\hat\alpha)$ with the same empirical content as the
standard quantum theory, and
for which no special set of surfaces is preferred.
As will be explained later, the model may be regarded
as the minimal realistic completion of the underlying quantum
theory, and the extension to a rather general class of local
lattice quantum theories is straightforward.
\vskip 5pt
The initial conditions are a spacelike surface
$\sigma_0$, the wavefunction $\Psi(\alpha |_{\sigma_0})$,
and a configuration $\hat\alpha|_0$ on $\sigma_0$
chosen according to the quantum mechanical probability distribution
(2), i.e. $|\Psi(\hat\alpha |_{\sigma_0})|^2$. (The dependence on $\sigma_0$
will be removed at the end of the construction by pushing it back
into the infinite past.) The evolution of the wavefunction
is taken to be the standard unitary one described before: thus in
the simplest case -- that of an elementary motion of the surface --
a single $R$-matrix is applied as in (1).
The specification of the dynamics is then completed by supplying a
rule for the evolution of the configuration variables $\hat\alpha_l$.
This is obtained very straightforwardly by imagining the initial
surface to advance stochastically by successive random elementary motions,
and requiring that given a particular motion, the joint distribution
of the $\hat\alpha_l$ on the new surface is always given by the
quantum mechanical result, (2). Making the simplest independence
assumption -- that the realizations $\hat\alpha_l$ are otherwise
random -- this prescription is enough to uniquely
define the model.
\footnote{$^2$}{For a related idea, see the discussion
of Bohmian theories in ref. 8 (and also 9), where it is suggested
that Bohm's guiding condition should be modified to act with respect
to a dynamically determined foliation of spacetime. Note
that here, where we do not have a guiding condition connecting the
configurations on adjacent disjoint surfaces, but only the weaker
constraint of consistency with (2), it will be essential that our
surfaces are generated by elementary motions and are more
``densely packed'' than a foliation.}
\vskip 5pt
To be precise, label the lines of the
lattice by $L$ and $R$ according to whether they represent
left-moving or right-moving null rays respectively.
With a given surface $\sigma_k$, we may then associate
a sequence $A|_k =(A_m)_{m=1}^{2N}$ of $L$'s and $R$'s
labelling the successive links it cuts as one moves from left to
right. Taking care to remember the periodicity in $m$,
the prescription for the
elementary motion to the next surface $\sigma_{k+1}$ is: pick
a $RL$ pair from this sequence at random and move the surface up
through the associated vertex, so that the pair is replaced by $LR$.
As in figure 1, let this motion be from the links $1$ and
$2$, to $1'$ and $2'$. To define the corresponding
evolution of the configuration on $\sigma_k$,
$\hat\alpha |_k$, we must specify the conditional
probability $f_\Psi(\hat\alpha |_{\sigma_k \to \sigma_{k+1}})$
of realizing the values $(\hat\alpha_{1'}, \hat\alpha_{2'})$ on the
newly-cut links, given all realized values $\hat\alpha_l$
up to that point. If we make the simplest possible assumption -- that
there is no conditional dependence on the realizations to the past of
$\sigma_{k+1}$ -- and apply the standard probability rule (2), then we
obtain the unique prescription

$$  f_\Psi(\hat\alpha |_{\sigma_k \to \sigma_{k+1}}) =
       {|\Psi(\alpha |_{\sigma_{k+1}})|^2 \over
   \sum_{\alpha_{1'}, \alpha_{2'}} |\Psi(\alpha |_{\sigma_{k+1}})|^2}
   \biggr\arrowvert_{\alpha |_{\sigma_{k+1}}=\hat\alpha |_{\sigma_{k+1}}}
     \eqno(5) $$

\noindent
where, by unitarity, the denominator can also be written as
$\sum_{\alpha_{1}, \alpha_{2}} |\Psi(\alpha |_{\sigma_{k}})|^2$.
\vskip 5pt
Together, the rules for $A|_k$ and $\hat\alpha|_k$ define a
discrete stochastic process $(A|_k, \hat\alpha |_k$) $(k=0,1,\ldots)$,
and (summing over the $A$'s) this generates a joint
probability distribution $p_{\Psi}(\hat\alpha|_{\sigma_0^+})$
in the $\hat\alpha_l$ over the entire lattice (on and) to the future
of the initial surface $\sigma_0$. Of course, the choice of a
particular initial surface should be regarded as an
artefact of the construction. The final step is thus to let $\sigma_0$
recede arbitrarily far into the past to obtain a distribution
$p_\Psi(\hat\alpha) = \lim_{\sigma_0 \to -\infty}
p_{\Psi}(\hat\alpha|_{\sigma_0^+})$ independent of any choice of
surfaces. The existence of this unique limiting distribution is shown
in the Appendix.
\vskip 5pt
The above is a dynamical description, in which
$p_\Psi (\hat\alpha)$ is regarded as being generated
by successive realizations on an advancing spacelike front.
It is also useful to think in a slightly different,
but equivalent way, in terms of a probabilistic
path integral. Consider a particular sequence of surfaces
$\gamma=(\sigma_k)$, and let
$p_{\Psi}^\gamma(\hat\alpha|_{\sigma_0^+})$ be the joint distribution
in all the $\hat\alpha_l$'s to the future of $\sigma_0$ conditional
on $\gamma$. By the above prescription we have

$$  p_{\Psi}^\gamma(\hat\alpha|_{\sigma_0^+}) =
      |\Psi(\hat\alpha |_{\sigma_0})|^2
   \prod_k f_\Psi(\hat\alpha |_{\sigma_k \to \sigma_{k+1}}).   \eqno(6)  $$

\noindent
Since, in the stochastic process, $\sigma_k$ evolves
by random elementary motions, all possible $\gamma$
contribute to the unconditional distribution
$p_\Psi(\hat\alpha |_{\sigma_0^+})$ with
equal weights. Thus,

$$  p_\Psi(\hat\alpha |_{\sigma_0^+}) \sim \sum_\gamma
          p_\Psi^\gamma(\hat\alpha|_{\sigma_0^+}),    \eqno(7)  $$

\noindent
or, letting $\sigma_0$ recede arbitrarily into the past,

$$  p_\Psi(\hat\alpha) \sim \sum_\gamma
          p_\Psi^\gamma(\hat\alpha),    \eqno(8)  $$

\noindent
where in each case, $\sim$ indicates the need for a normalization
constant. If the $\hat\alpha_l$'s are thought of as sources,
then (8) may be regarded as a sort of path integral, but involving
probability weights rather than amplitudes.
\vskip 5pt
The distribution $p_\Psi(\hat\alpha)$ is most easily understood
through the distributions $p_\Psi^\gamma(\hat\alpha)$ conditional
on $\gamma$. Each $p_\Psi^\gamma(\hat\alpha)$ is straightforward
to analyse since, {\it by construction},
its marginal distributions on all the $\sigma_k \in \gamma$
are just the quantum mechanical ones -- i.e.

$$ \hbox{\rm for each} \,\, \gamma: \qquad
\sum_{\hat\alpha |_{{\sigma_k}^c}} p_\Psi^\gamma(\hat\alpha) =
        |\Psi(\hat\alpha |_{\sigma_k})|^2
                  \qquad \forall \sigma_k \in \gamma.  \eqno(9) $$

\noindent
If it can be shown that some prediction follows from
$p_\Psi^\gamma(\hat\alpha)$, independently of choice of $\gamma$,
then it is also true of the distribution $p_\Psi(\hat\alpha)$.
\vskip 10pt
\centerline{\bf 5. Some properties of the model}
\vskip 5pt
The constraint (9) satisfied by the $p_\Psi^\gamma(\hat\alpha)$
is extremely powerful. It ensures
that the model has the same predictive content as the standard
quantum theory. In particular, it is straightforward to verify that
the model satisfies the conditions (I) and (II) which arose in our
earlier heuristic discussion.
\vskip 5pt
Condition (I) follows from the fact that for every $\gamma$, each
pair of ingoing links at a vertex
(and, indeed, each outgoing pair too) always lies on some
$\sigma_k \in \gamma$.
Thus the local marginal distributions of $p_\Psi(\hat\alpha)$
associated with these pairs
are just the quantum mechanical ones. It is worth remarking
that by virtue of this, they automatically satisfy locality --
that is to say, each such distribution is independent of the
$R$-matrices at spacelike separated points. This may be seen
explicitly by recalling that, as a result of local unitarity, such a
distribution may be written as the
appropriate marginal distribution of $|\Psi (\alpha|_\sigma)|^2$
for {\it any} $\sigma$ cutting the relevant pair;
pushing $\sigma$ back in time as far as possible, all $R$-matrices
spacelike separated from the pair will then lie in $\sigma$'s future,
thus making manifest their irrelevance to the pair's distribution.
\vskip 5pt
To understand how the model satisfies the continuity condition (II),
the crucial point to note is that the surface $\sigma_k$ evolves
by {\it local} (i.e. elementary)  motions: given an appropriate
conservation law, this automatically produces continuous
trajectories. Thus suppose that, as in figure 2(b), the $R$-matrices
conserve occupation number, and that the state $\Psi$ is an eigenstate
of the total occupation number (sum of $\alpha_l$'s on an arbitrary
surface). Then by (5), or equally (9), the {\it realized} occupation
number will be conserved through an elementary motion and
the stochastic process will generate $\hat\alpha$-configurations
that are continuous on the lattice. Thinking of
$\hat\alpha_l=1$ on a link as the presence of a ``particle'',
any given realization will thus be a set of continuous ``particle
paths'' through the lattice. Moreover, the behaviour of these paths
will be straightforward to analyse, since for any $\gamma$,
they will be cut by a ``dense'' sequence of
surfaces $\sigma_k \in \gamma$, on which, by (9), the
standard quantum prescription may be applied.
\vskip 5pt
In the particular case of the lattice massive Thirring model $\hat\alpha_l=1$
corresponds to the presence of a ``bare fermion''.
Going towards the continuum limit, and considering the non-relativistic
regime, the continuity property will carry over to the paths of
the physical particles as well.\footnote{$^3$}{To be more precise,
the (physical) particle position will only be well-defined on scales
greater than the Compton wavelength. The particle
paths will be obtained by spacetime coarse-graining
the $\hat\alpha_l$'s over this length scale and subtracting the
corresponding coarse-grained density in the vacuum state.}
If, for instance, a single particle is described by a
localized wave packet in the standard fashion, then the preceding
comments imply that in the stochastic model there will be a realized
particle trajectory which follows the motion of this packet.
Should the packet divide into several smaller packets,
the trajectory will follow one of them with a probability given
by the standard quantum mechanical probability of finding the particle
in that particular branch. The motion of a many-particle system will
be guided in the same way.
\vskip 10pt
\centerline{\bf 6. Measurement}
\vskip 5pt
To help to clarify these points and to make explicit the operational
equivalence of the scheme with the standard quantum theory it is
useful to consider a simple model of a measurement.
Let us introduce then a second set of
``apparatus'' variables $\beta_l$ on the links, again with the values
zero and one. The $R$-matrices will now be 8-dimensional and in
the state $\Psi$ the stochastic model will generate a distribution
$p_\Psi(\hat\alpha,\hat\beta)$.
\vskip 5pt
Suppose first that the $\beta_l$ are completely
independent of the $\alpha_l$, i.e. that all the $R$-matrices of the
complete system factorise as $R=R_\alpha\otimes R_\beta$, using an
obvious notation. At
every vertex let the only non-zero $\beta$-amplitudes be those depicted
in figure 2(b), but with the last two amplitudes which allow for
a change of direction now also set to zero. Most simply, for
instance, we may assume that at each vertex

$$ R_{\beta_1\,\beta_2}^{\beta_{1'}\beta_{2'}}=
     \delta_{\beta_1\beta_{2'}}\delta_{\beta_2\beta_{1'}}. \eqno(10)$$

\noindent Then, if the system is described by
a joint eigenstate of the $\beta$-occupation number operators on some
surface, the stochastic rule will generate a deterministic
evolution in the $\hat\beta_l$ consisting of a  set of
``$\hat\beta$-particles'' moving along null rays, independently
of the realizations $\hat\alpha_l$.
\vskip 5pt
Now suppose
we wish to ``measure'' $\hat\alpha_l$
on a particular link. We modify the amplitudes for the joint
system at the vertex $v$ from which this link is outgoing, so that

$$ R_{\alpha_1\,\alpha_2 \,\, 0 \,\, 0}^
      {\alpha_{1'}\alpha_{2'}\beta_{1'}\beta_{2'}}
         =R_{\alpha_1\,\alpha_2}^{\alpha_{1'}\alpha_{2'}}
          \delta_{\alpha_{1'}\beta_{1'}}\delta_{\alpha_{2'}\beta_{2'}},
                                  \eqno(11)  $$

\noindent and choose the other matrix elements at $v$,
$R_{\alpha_1\,\alpha_2\,\beta_1\,\beta_2}^
  {\alpha_{1'}\alpha_{2'}\beta_{1'}\beta_{2'}}$, to be consistent
with unitarity. Let the system be in the state corresponding to all
$\beta$-occupation numbers being zero on a surface $\sigma$ prior to $v$,
that is, again using an obvious notation,
$|\Psi \rangle_{\alpha} \otimes |0 \rangle_{\beta|_\sigma}$.
Then all the $\beta_l$ will realize the value zero except possibly
on the two null rays beginning at $v$, and on these,
the $\beta$-variables will realize the value one if and only if the
corresponding variables $\alpha_{1'}$, $\alpha_{2'}$ do. To talk
picturesquely, the realization of $\hat\alpha_l=1$ on one of
these links causes the emission of a $\hat\beta$-ray along the future
null extension of that link. This is our model of a measurement.
\vskip 5pt
We now use the apparatus to determine the joint distribution
of the $\hat\alpha$-variables at two spacelike separated links $l$ and $l'$.
We shall show that the result obtained is necessarily equal
to the standard quantum mechanical prediction; the extension to
an arbitrary number of variables is then straightforward.
We set up two measuring devices at the appropriate
vertices, and suppose furthermore that the two links are
inward-pointing, so that if $\hat\beta$-rays are produced,
they will intersect at some point $x$ in the future (see figure 4).
(Note that this is precisely the situation we argued was necessary
for the joint distribution to be given operational meaning; the
$\hat\beta$-rays are functioning here as the relevant records.)
By construction of the apparatus, the joint probability
for the $\hat\alpha$-variables at the chosen links,
$p(\hat\alpha_l,\hat\alpha_{l'})$, is equal to the joint probability
for the production of $\hat\beta$-rays, and this in turn
may be obtained from the marginal distribution for the $\hat\beta_l$
on any surface intersecting the causal past of the point $x$.
But for any $\gamma$ there will at least one such surface belonging
to $\gamma$, and so by (9) the agreement of the measured distribution
with the quantum mechanical prediction then follows immediately.
(Alternatively, simply apply (I) to the $\hat\beta_l$
on the pair of ingoing links at $x$.)

$$\displaylines{
 \epsfbox{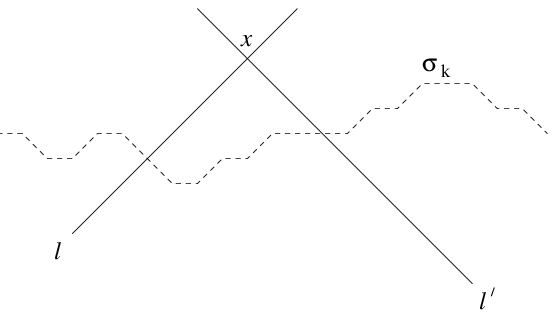} \cr}$$
\centerline{
 \vbox{\hsize=5 true in \noindent {\bf Figure 4.} {\it Analysis of a
  measurement. The solid lines denote the $\hat\beta$-rays
  produced by the realizations $\hat\alpha_l=1$ and $\hat\alpha_{l'}=1$.
  The dashed line is a surface $\sigma_k$ picked out by a particular
  realization of the stochastic process.}}}

\vskip 5pt
In fact, it should be clear that the same result holds rather
more generally. It is not necessary that the records actually be
brought together to the same point -- only that, for every $\gamma$,
there is a surface $\sigma_k \in \gamma$ cutting them, so that
(9) may then be invoked. This circumstance will obtain
provided that the records are sufficiently persistent. On the
other hand, if it does not -- as might happen for instance
if a record is prematurely destroyed -- then
{\it a fortiori} there is some surface that separates the
records, contact between them is excluded,
and the joint distribution is deprived of operational significance.
Harmony with the standard predictions is thus always maintained.
Exactly the same considerations apply immediately to the measurement of
the joint distribution $p(\hat\alpha_{l_1},\ldots , \hat\alpha_{l_n})$
of an arbitrary set of spacelike separated variables.
\vskip 5pt
Of course the simplicity of the field theory and the demands of
explicitness mean that our model
of the measuring apparatus is rather crude. In practice,
the appropriate amplitude structure leading to the
continuous, quasi-deterministic trajectories of the classical
regime will generally emerge from the quantum mechanical evolution
law for systems with large numbers of degrees of freedom (decoherence)
rather then being imposed by fiat at the microscopic level as above.
The principle, however, remains precisely the same.
\vskip 10pt
\centerline{\bf 7. Remarks}
\vskip 5pt
As we have already observed, the local (two-variable) marginal
distributions satisfy locality. By contrast, a marginal
distribution associated with a more extended region of spacetime will
generally depend on the $R$-matrices at points spacelike separated from
that region. This is a rather strong form of the non-locality
that we expected on general grounds from the outset. Nonetheless, because
of the operational equivalence of the model with standard quantum theory,
pathologies such as superluminal signalling are avoided,
and the ``peaceful coexistence''$^{(17)}$ of quantum theory and
special relativity is left undisturbed. It is perhaps interesting
to note that models of the second type (in which the unitary
evolution is stochastically modified) are able to enjoy a
better locality property in which the
marginal distribution of events associated with a region is
completely independent of the evolution law in spacelike separated
regions (though of course still non-locally correlated with other
events in those regions).$^{(18,19)}$ The stronger non-locality we have
found here is presumably the consequence of adhering
to a strictly unitary evolution law for the wavefunction.
\vskip 5pt
In spite of this non-locality, it is worth emphasizing that the
relativistic setting is crucial for the viability of the model.
Were the underlying quantum theory non-relativistic, then the only
allowed ``elementary motion'' would be between adjacent constant time
slices, and the rule (5) would reduce simply to (2),
with variables on different slices being independent. The model would
thus collapse to Bell's version of Everett$^{(16)}$, complete with its
pathological absence of sensible histories. Indeed, one may think
of the model as simply this Bell-Everett
prescription, but with its two defects -- absence of sensible
histories and frame-dependence -- simultaneously cured by the use
of the random sequence of surfaces that the locality of the
evolution law of the quantum theory makes possible.
\vskip 5pt
Another, rather suggestive way of thinking about the model is in
terms of a stochastic accumulation of events on
an advancing spacelike front -- a picture somewhat like that proposed
by Haag$^{(20)}$. To make this more precise, define
an ``event'' as the pair of realizations on the outgoing links
at a particular vertex. The sum (8) is over all
possible total orderings of events consistent
with the partial ordering defined by the causal structure, so that
an event may occur only after all the events in its causal past.
Given such an ordering, the rule (5) simply describes the conditional
probability of the realizations $\hat\alpha_l$ associated with an event,
given all the events up to that point. Furthermore, (5) is the simplest
assumption compatible with standard quantum mechanics,
since it assumes a conditional dependence only on the other
events of the current spacelike boundary, $\sigma_{k+1}$; any other
rule would involve an additional dependence on the events to the past of
this boundary, and so require an extension of the basic prescription (2).
In this sense, the model is the minimal realistic completion of the
underlying quantum theory.
\vskip 5pt
It is tempting to speculate that by
examining the distribution of events sufficiently far back in time,
one could infer the state $\Psi$ to arbitrary accuracy. One could
then regard the quantum state and unitary evolution as purely auxiliary
concepts, and think of the stochastic law governing the evolution
as being one in which the probability of ``the next event'' depends
simply on all the events which have ever preceded it (see also
refs. 21 and 11 for related points).
\vskip 10pt
\centerline{\bf 8. Generalization}
\vskip 5pt
To conclude, we note that in constructing the stochastic
model we have only made use of the causal
structure and the local unitary evolution law of the underlying lattice
field theory, together with a particular application of the standard
probability interpretation of the wavefunction. There is therefore
an immediate generalization to a broad class of such theories.
\vskip 5pt
Consider a locally finite, partially ordered set
of points, $\cal P$, with the partial ordering $x<y$
describing the causal relation ``$y$ is in the future of $x$'', and
take for a lattice the corresponding Hasse diagram.
(See e.g. Stanley$^{(22)}$ for definitions, but note that we are using
the word ``lattice'' in a non-technical sense.) To each link $l$ assign
a number $n_l$ of states, labelled by $\alpha_l \, (=0,\ldots,n_l-1)$ say,
and suppose, as a condition on $\cal P$, that
this can be done so that, at each vertex, the number of ingoing states
equals the number of outgoing ones.
By associating with the vertices unitary matrices connecting these
states, one then obtains a local quantum theory on the lattice.
\vskip 5pt
In this general framework, a spacelike surface $\sigma$ is
a cut through the links that intersects each
everywhere future-directed path exactly once.
At a vertex all of whose ingoing links are cut by $\sigma$,
an elementary motion can be performed by moving
the cut to the outgoing ones.
One may thus associate a stochastic
process with the quantum theory in exactly the same way
as before -- namely, by allowing a surface to evolve by random
elementary motions, and taking the probabilistic law for the
realizations on the newly cut links to be the obvious
generalization of (5) -- and again as before, this process may
be used to generate a joint probability distribution
$p_\Psi(\hat\alpha)$ for the realizations $\hat\alpha_l$ over the
lattice.
\vskip 5pt
In principle it should be possible to construct a similar process
for any local quantum theory whose evolution law can be regularized
on a suitable causal lattice. In the case of a gauge
theory, for instance, this will presumably involve group-valued
variables on the links and a local evolution law associated with
plaquettes rather than vertices. An important further question of
course is whether one can obtain well-defined continuum limit.
Needless to say, a formulation that employed continuum concepts
from the outset, to be regularized in a convenient way at a second
stage, would be highly desirable.
\vskip 20pt
\centerline{\bf Appendix}
\vskip 5pt
To show the existence of the distribution
$p_\Psi(\hat\alpha)$ in the limit that $\sigma_0$ is
pushed back arbitrarily into the past, it is useful to introduce
the idea of the ``time'' $t_\sigma$ associated with a
surface $\sigma$. The flat surfaces $\sigma_t$ (see figure 1)
define a time coordinate $t$, whose units we can choose so
that the time between adjacent surfaces is
${1\over2}$. For a general surface $\sigma$, we then define
$t_\sigma$ to be the average of the times of each of its links.
Note that the sequence $(A_m)_{m=1}^{2N}$
associated with a surface $\sigma$ defines its time up to an integer
$T(=[t_\sigma])$; thus any surface is uniquely specified by the
data $((A_m), T)$.
\vskip 5pt
The random evolution law for $\sigma_k$ corresponds to a homogeneous
Markov process $(A |_k)$ $(k=0,1,\ldots)$: at each step an
$RL$ pair is chosen at random and replaced by $LR$.
Moreover, the process is finite and irreducible.
Consequently there is a unique stationary distribution.
It follows that, as $t_{\sigma_0} \to -\infty$, the
distribution over the space of sequences of surfaces between
arbitrary finite times tends to a unique limiting distribution.
Furthermore, {\it given} a particular sequence $\gamma=(\sigma_k)$,
we have: (i) the sequence of realizations $(\hat\alpha |_k)$ is
also a Markov process, since the conditional probability of
$\hat\alpha |_{k+1}$ given all previous realizations depends only
on $\hat\alpha |_k$, through (5); and (ii) the absolute probability
of $\hat\alpha |_k$ is always $|\Psi(\hat\alpha |_{\sigma_k})|^2$.
It is then immediate that, in any finite region of the lattice $D$, there
is also a unique limiting distribution in the $\hat\alpha_l$. To
generate it, simply take any time $t$ for which all $\sigma$ with
$t_\sigma=t$ lie to the past of $D$, and use as initial conditions
the distributions $|\Psi(\hat\alpha |_\sigma)|^2$ on each such
$\sigma$, weighted according to the stationary distribution
for the surfaces.
\vskip 5pt
(For completeness note that the periodicity of the lattice
means that any allowed $(A_m)_{m=1}^{2N}$ must
have an equal number of $R$'s and $L$'s, and so the Markov process
$(A|_k)$ is over a space of $(2N)!/N!N!$ states.
It is also straightforward to show that each
state has period $2N$, and appears in the the stationary distribution
with a weight given by the number of $RL$ pairs it contains.)
\vskip 20pt
\centerline{\bf Acknowledgements}
\vskip 5pt
I am very grateful to Adrian Kent for many useful discussions,
and for a critical reading of the manuscript. I am also grateful
for conversations with Fay Dowker, Klaas Landsman, Geof Nicholls
and G\'erard Watts, and thank Rafael Sorkin for a helpful criticism,
and Sheldon Goldstein for helpful comments, criticisms and discussion.
\vskip 30pt
\centerline{\bf References}
\vskip 5pt \noindent
\item{1.} J.S. Bell, {\it Speakable and unspeakable in
quantum mechanics.} (Cambridge University Press, Cambridge 1987).
\vskip 5pt \noindent
\item{2.} D. Bohm, A suggested interpretation of the quantum theory in
terms of ``hidden'' variables I \& II, {\it Phys. Rev.} {\bf 85}
(1952) 166-193.
\vskip 5pt \noindent
\item{3.} J.S. Bell, Beables for quantum field theory, preprint CERN-TH
4035/84 (1984), in ref. 1, pp. 173-180.
\vskip 5pt \noindent
\item{4.} S.M. Roy \& V. Singh, Generalized beable quantum field theory,
{\it Phys. Lett.} B {\bf 234} (1990) 117-120.
\vskip 5pt \noindent
\item{5.} P. Pearle, Ways to describe state vector reduction,
{\it Phys. Rev.} A {\bf 48} (1993) 913-923.
\vskip 5pt \noindent
\item{6.} I.C. Percival, Primary state diffusion, {\it Proc. Roy. Soc.} A
{\bf 447} (1994) 189-209.
\vskip 5pt \noindent
\item{7.} J.F. Clauser \& A. Shimony, Bell's theorem: experimental
tests and implications, {\it Rep. Prog. Phys.} {\bf 41} (1978) 1881-1927.
\vskip 5pt \noindent
\item{8.} D. D\"urr, S. Goldstein \& N. Zhang\'\i, On a realistic theory
for quantum physics, in {\it Stochastic processes, physics and
geometry}, S. Albeverio et al., eds. (World Scientific, Singapore 1990),
pp. 374-391.
\vskip 5pt \noindent
\item{9.} K.S. Berndl, D. D\"urr, S. Goldstein \& N. Zhang\'\i, Towards
a relativistic quantum theory of particles (having trajectories).
Talk given at Third UK
Conference on Foundations of Quantum Theory and Relativity: Cambridge,
September 15th (1994).
\vskip 5pt \noindent
\item{10.} K.S. Berndl \& S. Goldstein, Comment on ``Quantum mechanics,
local realistic theories, and Lorentz-invariant realistic theories'',
{\it Phys. Rev. Lett.} {\bf 72} (1994) 780.
\vskip 5pt \noindent
\item{11.} D. D\"urr, S. Goldstein \& N. Zhang\'\i, Quantum equilibrium
and the origin of absolute
uncertainty, {\it J. Stat. Phys.} {\bf 67} (1992) 843-907.
\vskip 5pt \noindent
\item{12.}G.C. Ghirardi, R. Grassi \& P. Pearle, Relativistic dynamical
reduction models: general framework and examples,
{\it Found. Phys.} {\bf 20} (1990) 1271-1316.
\vskip 5pt \noindent
\item{13.} C. Destri, C. \& H.J. de Vega, Light-cone lattice approach to
fermionic theories in 2D,
{\it Nucl. Phys.} B {\bf 290} (1987) 363-391.
\vskip 5pt \noindent
\item{14.} A. Fine, Joint distributions, quantum correlations, and commuting
observables, {\it J. Math. Phys.} {\bf 23} (1982) 1306-1310.
\vskip 5pt \noindent
\item{15.} L. Hardy, Quantum mechanics,
local realistic theories, and Lorentz-invariant realistic theories,
{\it Phys. Rev. Lett.} {\bf 68} (1992) 2981-2984.
\vskip 5pt \noindent
\item{16.} J.S. Bell, Quantum mechanics for cosmologists, in {\it Quantum
Gravity 2}, C. Isham, R. Penrose \& D. Sciama, eds. (Clarendon Press,
Oxford, 1981). Reprinted in ref 1, pp. 117-138.
\vskip 5pt \noindent
\item{17.} A. Shimony, Events and processes in the quantum world,
in {\it Quantum concepts in
space and time}, R. Penrose \& C.J. Isham, eds.
Oxford University Press, New York 1986), pp. 182-203.
\vskip 5pt \noindent
\item{18.} J.S. Bell, Are there quantum jumps? in {\it Schr\"odinger: Centenary
celebration of a polymath}, C.W. Kilmister, ed. (Cambridge University
Press, Cambridge, 1987) pp. 41-52. Reprinted in ref 1, pp. 201-212.
\vskip 5pt \noindent
\item{19.} N. Gisin, Stochastic quantum dynamics and relativity,
{\it Helv. Phys. Act.} {\bf 62} (1989) 363-371.
\vskip 5pt \noindent
\item{20.} R. Haag, Fundamental irreversibility and the concept of events,
{\it Comm. Math. Phys.} {\bf 132} (1990) 245-251.
\vskip 5pt \noindent
\item{21.} A. Kent, ``Quantum jumps'' and indistinguishability,
{\it Mod. Phys. Lett.} {\bf 4} (1989) 1834-45.
\vskip 5pt \noindent
\item{22.} R.P. Stanley, {\it Enumerative Combinatorics},
(Wadsworth \& Brooks/Cole Advanced Books, Monterey 1986).
\bye